\documentclass[12pt]{iopart}

\usepackage{amssymb}
\usepackage{graphicx}
\usepackage{epsfig,multicol,subfigure}
\usepackage{dcolumn,color}

\begin{document}

\title{Matching the linear spectra of twinlike defects}

\author{Yuan Zhong$^{1,2}$, Yu-Xiao Liu$^{1}$\footnote{Corresponding author.}}

\address{$^{1}$Institute of Theoretical Physics, Lanzhou University,
           Lanzhou 730000, People's Republic of China \\
         $^{2}$IFAE, Universitat Aut$\grave{o}$noma de Barcelona, 08193 Bellaterra, Barcelona, Spain}

\ead{zhongy2009@lzu.edu.cn, liuyx@lzu.edu.cn}

%\begin{abstract}
%Twinlike defects refer to topological defect solutions of some apparently different field models that share the same defect configuration and the same energy density. Usually, one can distinguish twinlike defects in terms of their linear spectra, but in some special cases twinlike defects even share the same linear spectrum. In a recent publication [C. Adam and J. Queiruga, Phys. Rev. D 85, 025019 (2012)], the authors investigated the algebraic conditions for two twinlike defects that have identical linear spectrum. In this paper, we reexamine these algebraic conditions from the viewpoint of the normal modes of the linear fluctuations. We obtain a simpler but less restricted algebraic condition. Our results open a new window for interesting models that violate the algebraic constraints in Adam-Queiruga's construction. We also extend our discussion to braneworld models, where gravity plays an important role.
%\end{abstract}

\begin{abstract}
Twinlike defects refer to topological defect solutions of some apparently different field models that share the same defect configuration and the same energy density. Usually, one can distinguish twinlike defects in terms of their linear spectra, but in some special cases twinlike defects even share the same linear spectrum. In this paper, we derive the algebraic conditions for two twinlike defects to share identical linear spectrum from the viewpoint of the normal modes of the linear fluctuations. We also extend our discussion to braneworld models, where gravity plays an important role.
\end{abstract}


%\PACS{
%      {11.10.Lm}{Nonlinear or nonlocal theories and models}   \and
%      {11.27.+d}{Extended classical solutions; cosmic strings, domain walls, texture}\and
%     {04.50.-h}{Higher-dimensional gravity and other theories of gravity}
%     }% end of  PACS codes
%}

%\maketitle
%
\section{Introduction}
Many field models were found to support classical solutions which have finite energies and extend field configurations. Such solutions are referred to as solitons or defects, and have been applied in many branches of physics research, ranging from condensed matter physics~\cite{BishopSchneider1978}, particle physics~\cite{RebbiSoliani1984}, to cosmology~\cite{VilenkinShellard2000}. The simplest defect solutions are the kinks, which are solutions of two-dimensional models with only a single real scalar field. The temptation for finding higher-dimensional defect solutions in canonical one-component scalar field theory is forbidden by the Derrick theorem~\cite{Derrick1964}. However, higher-dimensional defect solutions were found in some scalar field models, where the scalar fields have noncanonical kinetic terms~\cite{Diaz-Alonso1983,Babichev2006,AvelinoBazeiaMenezes2011}. Such noncanonical scalar fields are called $K$-fields, which were originally introduced in cosmology to trigger the cosmological inflation~\cite{Armendariz-PiconDamourMukhanov1999,GarrigaMukhanov1999,Armendariz-PiconMukhanovSteinhardt2001,Armendariz-PiconMukhanovSteinhardt2000,Scherrer2004}. Now $K$-fields have been repeatedly studied in string theories~\cite{Sen2002,Sen2003,Sen2005}, braneworld models~\cite{AdamGrandiSanchez-GuillenWereszczynski2008,BazeiaGomesLosanoMenezes2009,LiuZhongYang2010,ZhongLiu2013,ZhongLiuZhao2014a}, and massive gravity theories~\cite{Hinterbichler2012,Rham2014}. It should be interesting to consider the existence and properties of various defect solutions in $K$-field models.

Recently, it was found that defect solutions of a standard scalar field model might have ``twins" in some $K$-field models. Two twinlike defects share the same field configuration and energy density. We call the corresponding models the twinlike models. The first couples of twinlike models were reported in Ref.~\cite{AndrewsLewandowskiTroddenWesley2010}, where the authors studied the domain wall solution of a scalar Dirac-Born-Infeld (DBI) type model\footnote{The original DBI model was proposed in 1934 by Born and Infeld~\cite{BornInfeld1934} to solve the problem of the divergence of electron's self-energy in electromagnetic field theory. In their model, the kinetic term of the electromagnetic field is written under a square root. In the 1980s, there was a revival in the study of DBI model as it was found to arise in the low-energy limit of string/brane physics~\cite{FradkinTseytlin1985,Leigh1989}. Nowadays, similar models have been considered in scalar~\cite{Gibbons2002,KimKimLee2003,Sen2003} and gravitational~\cite{BanadosFerreira2010} systems, sometimes these models are also called DBI models for simplicity.} in four-dimensional flat space-time.

Suppose the scalar field possesses the following Lagrangian:
\begin{equation}
\label{Lgen}
\mathcal{L}=\mathcal{L}(X,\phi),
\end{equation}
where $X=-\frac12\partial_\mu\phi\partial^\mu\phi$ is the kinetic term of $\phi$, and $\mu=0,1,2,3$ is the space-time indices. The authors of Ref.~\cite{AndrewsLewandowskiTroddenWesley2010} studied a special case where
\begin{equation}
\label{LDBI}
\mathcal{L}=\mathcal{L}_{\textrm{DBI}}=1-(1+U(\phi))\sqrt{1-2X},
\end{equation}
where $U(\phi)$ is a function of $\phi$.
The authors of Ref.~\cite{AndrewsLewandowskiTroddenWesley2010} found that when
\begin{equation}
\label{potential}
 U=\sqrt{1-2V(\phi)}-1,
\end{equation}
no matter what the form $V$ takes, the model (\ref{LDBI}) always possesses a solution ( dubbed as ``doppelg\"anger domain wall"), which has the same field configuration and energy density as the wall solution of the canonical model $\mathcal{L}_0=X-V(\phi)$.

Thus, by definition the DBI model specified by the Lagrangian (\ref{LDBI}) and the potential (\ref{potential}) is a twinlike model of the canonical model. But the DBI model is merely one of the infinite twinlike models of the canonical model. Suppose the solution of an arbitrary noncanonical model $\mathcal{L}(X,\phi)$ traces out a curve $C$ on the $(X,\phi)$ plane. Then the sufficient and necessary conditions for this model to be a twin of the canonical one are~\cite{AndrewsLewandowskiTroddenWesley2010}:
\begin{eqnarray}
\label{cre1}
  \mathcal{L} &=& \mathcal{L}_0,~~~~\, \textrm{on} \quad C, \\
  \label{cre2}
  \mathcal{L}_{,X} &=& \mathcal{L}_{0,X},~~ \textrm{on} \quad C.
\end{eqnarray}
Here and in what follows we always use shortcuts like $\mathcal{L}_{,X}\equiv \frac{\partial\mathcal{L}}{\partial X}$, \emph{et al.}.
These criteria do not uniquely determine the form of $\mathcal{L}$. So there are infinite twinlike models for the canonical model in the case with a single scalar field and without gravity.

For models with multi-scalar fields or with gravity, the above criteria are no longer valid. Nevertheless, it is still possible to construct twinlike models of braneworld models~\cite{BazeiaDantasGomesLosanoMenezes2011}, cosmological models~\cite{BazeiaDantas2012}, compacton models~\cite{BazeiaLobMenezes2012}, multi-scalar field models~\cite{BazeiaLobaoLosanoMenezes2014}, and self-dual Abelian-Higgs theories~\cite{BazeiaHoraMenezes2012}.

According to Ref.~\cite{AndrewsLewandowskiTroddenWesley2010}, twinlike defects usually have different linear spectra, and thus can be distinguished by analyzing their linear fluctuations. It is natural to ask if it is possible that two twinlike defects also share the same linear spectrum? In Ref.~\cite{BazeiaMenezes2011}, Bazeia and Menezes gave us a positive answer by providing the first example of twinlike models that support twinlike defects with identical linear spectrum. For simplicity, let us call twinlike defects with identical linear structure \emph{the special twinlike defects} and call the corresponding models \emph{the special twinlike models} from now on.

A number of special twinlike models were constructed later in Ref.~\cite{AdamQueiruga2012}, where the Lagrangian $\mathcal{L}=\mathcal{L}(X,V)$ was assumed to be a function of $X$ and $V$ (rather than a function of $X$ and $\phi$). With this Lagrangian, the criteria (\ref{cre1})-(\ref{cre2}) can be rewritten as
 \begin{eqnarray}
 \label{ag1}
  \mathcal{L}| = -2 V, \\
   \label{ag2}
  \mathcal{L}_{,X}| = 1,
\end{eqnarray}
where the vertical line $|$ represents taking the on-shell condition $X=-V$ (see also Ref.~\cite{AdamQueiruga2011}).
If in addition to Eqs.~(\ref{ag1})-(\ref{ag2}), the Lagrangian $\mathcal{L}$ also satisfies the following equations~\cite{AdamQueiruga2012}:
\begin{eqnarray}
\label{crespe1}
\mathcal{L}_{,XX}|=0, \\
\label{crespe2}
\left[ \mathcal{L}_{,XV}+2V(\mathcal{L}_{,XXX}-\mathcal{L}_{,XXV})\right] | = 0, \\
\label{crespe3}
\left[ \mathcal{L}_{,VV}+\mathcal{L}_{,XV}+2V(\mathcal{L}_{,XVV}-\mathcal{L}_{,XXV}) \right] | =0, \\
\label{crespe4}
(\mathcal{L}_{,V}+2V\mathcal{L}_{,XV})| =-1,
\end{eqnarray}
then the defect solution of the model would share the same defect configuration, energy density, and linear spectrum with the canonical defect.

Note that Eqs.~(\ref{crespe1})-(\ref{crespe4}) were obtained by comparing the noncanonical linear perturbation equation with the canonical one. However, we notice that  the authors of Ref.~\cite{AdamQueiruga2012} did not simplify the former to the final form. In fact, in one of our recent work~\cite{ZhongLiu2014}, we have shown that the linear spectrum of a noncanonical model depends only on the form of $\mathcal{L}_{,X}$ and $\mathcal{L}_{,XX}$. The term $\mathcal{L}_{,\phi}$ or equivalently $\mathcal{L}_{,V}$ can be replaced in terms of $X$, $\mathcal{L}_{,X}$, $\mathcal{L}_{,XX}$, and their derivatives. Besides, for twinlike defects, $\mathcal{L}_{,X}$ is constrained by Eq.~(\ref{ag2}), we expect that the linear spectrum is determined only by $\mathcal{L}_{,XX}$. Therefore, instead of Eqs.~(\ref{crespe1})-(\ref{crespe4}), we need only one equation of $\mathcal{L}_{,XX}|$ to tell if a noncanonical model is the special twin of the canonical one.

As we will show below that Eq.~(\ref{crespe1}) is merely a sufficient condition for a noncanonical model to be the special twin of the canonical one, and that some noncanonical models despite violate Eq.~(\ref{crespe1}) can still be special twinlike models of the canonical one. Aside from this, there is no reports on special twinlike models in gravitational systems. Thus, the aims of this paper are twofold: to derive the most general criterion for special twinlike models in two-dimensional flat space-time, and to generalize this criterion to the braneworld model, which is a simple gravitational model of current interesting in high energy physics, and whose linear structure is well known.

In the next section, we consider special twinlike models in two-dimensional flat space-time. We show that Eqs.~(\ref{crespe1})-(\ref{crespe4}) can be replaced by a single equation. Using this equation as well as Eqs.~(\ref{ag1})-(\ref{ag2}), we construct two special twinlike models (with nontrivial $\mathcal{L}_{,XX}|\neq 0$) for the canonical model. Then in section~\ref{sec3}, we extend our discussions to braneworld models. We first derive the equation for the normal mode of the linear fluctuations. From this equation we can read out the constraint for the Lagrangian of the special twinlike braneworld models. Two example models are constructed.
Our results will be summarized in section~\ref{sec4}.

\section{Special twinlike defects in two-dimensional flat space-time}
\label{sec2}

In two-dimensional flat space-time ($x^0=t$, $x^1=x$, $\eta^{\mu\nu}=$diag$(-1,1)$), the standard kinetic term for a static scalar field $\phi=\phi(x)$ is
\begin{eqnarray}
X=-\frac12\phi'^2.
\end{eqnarray}
In this section, a prime represents the derivative with respect to $x$.
Consider a model described by the Lagrangian
\begin{eqnarray}
\label{action}
\mathcal{L}=\mathcal{L}(X,\phi),
\end{eqnarray}
we obtain the following equation of motion
\begin{eqnarray}
\label{EoM}
-\mathcal{L}_{,\phi}=(\mathcal{L}_{,X} \phi')'.
\end{eqnarray}
One can easily integrate Eq.~(\ref{EoM}) to obtain the following equation~\cite{AdamQueiruga2011}
\begin{eqnarray}
\label{eom2D}
\mathcal{L}-2X\mathcal{L}_{,X}=0.
\end{eqnarray}
The energy density (the Hamiltonian density) is simply
\begin{eqnarray}
\label{2dlag}
\rho=-\mathcal{L}.
\end{eqnarray}

For the standard model $\mathcal{L}_0=X-V$, Eq.~(\ref{eom2D}) reduces to
\begin{eqnarray}
\label{eq2d}
X=-V.
\end{eqnarray}
This is a first-order differential equation for $\phi(x)$. Specifying the form of $V(\phi)$, one would obtain the solution of $\phi(x)$. By substitute Eq.~(\ref{eq2d}) into Eq.~(\ref{2dlag}), we get the energy density for the canonical defect
\begin{eqnarray}
\rho_0=2V.
\end{eqnarray}

To construct twinlike defect models for the canonical model, it is convenient to rewrite $\mathcal{L}(X,\phi)$ as
\begin{equation}
\mathcal{L}=\mathcal{L}(X,V).
\end{equation}
In order for the noncanonical defects to have the same configuration as the canonical defect, we require $X=-V$ as the on-shell equation. That means no matter how complicate a Lagrangian is, the final equation of motion must be $X=-V$.

To ensure that the noncanonical defects share the same energy density with the canonical defect, we require
\begin{equation}
\rho=-\mathcal{L}|=\rho_0=2V.
\end{equation}
This is nothing but Eq.~(\ref{ag1}). Using Eq.~(\ref{eom2D}), we immediately obtain Eq.~(\ref{ag2}):
\begin{equation}
\mathcal{L}_{,X}|= 1.
\end{equation}

A noncanonical model whose Lagrangian satisfies Eqs.~(\ref{ag1})-(\ref{ag2}) must be a twinlike model of the canonical model. Let us cite here the twinlike models constructed by Adam and Queiruga in Ref.~\cite{AdamQueiruga2012}:
\begin{enumerate}
  \item Model 1:
\begin{eqnarray}
\mathcal{L}^{\textrm{mod-1}}=\sum_{i=3,5,\cdots}^{2N+1}f_i(V)(X+V)^i
+X-V,
\end{eqnarray} where $f_i(V)\geq 0$ are arbitrary functions of $V$.
  \item Model 2:
  \begin{eqnarray}
\mathcal{L}^{\textrm{mod-2}}&=&1-\sqrt{1+2V}\sqrt{1-2X} %\nonumber\\
 -\frac{(X+V)^2}{2 (1+2 V)}.
\end{eqnarray}
\end{enumerate}
One can easily proof that both of the above models satisfy Eqs.~(\ref{ag1})-(\ref{ag2}), and therefore are twinlike models of the canonical model. In fact, in addition to Eqs.~(\ref{ag1})-(\ref{ag2}), the above models also satisfy Eqs.~(\ref{crespe1})-(\ref{crespe4}). So, models 1 and 2 are special twinlike models of the canonical model.

As mentioned in the introduction, our aim is to construct special twinlike models that has nontrivial $\mathcal{L}_{,XX}|$. To do this, we need to investigate the structure of the linear spectrum of the general noncanonical model.

\subsection{The quadratic action and the linear spectrum}
\label{secQuadratic}
The linearization of the model (\ref{action}) has been conducted in Ref.~\cite{ZhongLiu2014}. Here we briefly review the results. Expanding the Lagrangian in Eq.~(\ref{action}) to the second-order of the field fluctuation $\delta\phi$, we obtain
\begin{eqnarray}
\label{Lm}
 {\delta ^{(2)}}{\cal L} &=& \frac{1}{2}\Big\{ {{\cal L}_{,\phi \phi }}{(\delta \phi )^2} + {{\cal L}_{,XX}}{(\phi ')^2}{(\delta \phi ')^2} \nonumber\\
  &-& {{\cal L}_{,X}}{\partial ^\mu}\delta \phi {\partial _\mu}\delta \phi  - 2{{\cal L}_{,\phi X}}\phi '\delta \phi \delta \phi ' \Big\} .
\end{eqnarray}
Using the equation of motion, one can eliminate $\mathcal{L}_{,X\phi}$ and $\mathcal{L}_{,\phi\phi}$. By defining $\mathcal{G}\equiv \delta \phi\sqrt{\mathcal{L}_{,X}}$, we obtain
\begin{equation}
 {\delta ^{(2)}}\mathcal{L} = \frac{1}2\left\{
   -\mathcal{G}\partial_t^2\mathcal{G} +U(x)\mathcal{G}^2
+\gamma\mathcal{G}\mathcal{G}''\right\},
\end{equation}
where
\begin{equation}
U(x)=-\gamma \frac{z''}{z}-\frac{z'}{z}\gamma '-\frac{1}{2}\gamma '',
\end{equation}
and
\begin{equation}
\label{zf}
z=\phi '\mathcal{L}_{,X}^{1/2},\quad
\gamma=1+2\frac{\mathcal{L}_{,XX} X}{\mathcal{L}_{,X}}.
\end{equation}
When $\gamma>0$, we can introduce a new coordinate $x^{\ast}$
\begin{equation}
\label{RWcoord}
\frac{dx^{\ast}}{dx}\equiv\gamma^{-1/2}
\end{equation} to rewrite the quadratic action as
\begin{eqnarray}
{\delta ^{(2)}}{S_\mathcal{G}} &= &\frac{1}{2}\int dtdx^{\ast}{\sqrt{\gamma}} %\nonumber\\
 %&\times &
 \left\{
   -\mathcal{G}\partial_t^2\mathcal{G} +U_{\textrm{eff}}(x^{\ast})\mathcal{G}^2
+ \mathcal{G} \ddot{\mathcal{G}}\right\},
\end{eqnarray}
where
\begin{eqnarray}
U_{\textrm{eff}}(x^{\ast})&\equiv & U(x^{\ast})+\frac{1}{4\sqrt\gamma}\frac{d}{dx^{\ast}}\left(\frac{ \dot\gamma}{ \sqrt{\gamma }}\right).
\end{eqnarray}
Here, an over dot represents the derivative with respect to $x^\ast$.

Obviously, the normal mode of the quadratic action is
\begin{equation}
\hat{\mathcal{G}}=\frac{1}{\sqrt{2}}\gamma^{1/4}\mathcal{G}.
\end{equation}
In terms of $\hat{\mathcal{G}}$, the quadratic action reads
\begin{equation}
\label{G2}
 {\delta ^{(2)}}{S_{\hat{\mathcal{G}}}} = \int dtdx^{\ast}\hat{\mathcal{G}}
 \Big \{-\partial_t^2\hat{\mathcal{G}}+ \ddot{\hat{\mathcal{G}}} -\frac{\ddot{\theta}}{\theta}\hat{\mathcal{G}}
\Big\},
\end{equation}
where
\begin{equation}
\theta \equiv\gamma^{1/4}z .
\end{equation}

From the quadratic action of $\hat{\mathcal{G}}$, we know that for
\begin{eqnarray}
\label{criterion}
\mathcal{L}_{,X}>0,\quad \gamma>0,
\end{eqnarray}
the linear perturbation satisfies a Schr\"odinger-like equation
\begin{equation}
\label{schroScalar}
-\ddot{\hat{\mathcal{G}}} +\frac{\ddot{\theta}}{\theta}\hat{\mathcal{G}}
=-\partial_t^2\hat{\mathcal{G}}.
\end{equation}

From the linear perturbation equation (\ref{schroScalar}), the linear spectrum of $\hat{\mathcal{G}}$ is determined only by the effective potential, and therefore, by $\theta$. For the standard model, $\gamma_0=1$ and $z_0=\phi'$, we have $\theta_0=\phi'$. Therefore, to obtain a special twinlike model which satisfies $\ddot{\theta}/\theta=\ddot{\theta}_0/\theta_0$, we require
\begin{equation}
\theta \propto \theta_0.
\end{equation}
Using the definitions in Eq.~(\ref{zf}) and $\mathcal{L}_{,X}|=1$, we immediately obtain
\begin{equation}
(\mathcal{L}_{,XX}X)|=c,
\end{equation}
or equivalently,
\begin{equation}
\label{creNew2D}
\mathcal{L}_{,XX}|=-\frac c{V},
\end{equation}
where $c$ is a positive constant.

\subsection{Explicit examples}
With Eq.~(\ref{creNew2D}), we can now construct a new class of special twinlike models that are essentially different from $\mathcal{L}^{\textrm{mod-1}}$ and $\mathcal{L}^{\textrm{mod-2}}$. But for comparison, we would like to construct our models by simply modify $\mathcal{L}^{\textrm{mod-1}}$ and $\mathcal{L}^{\textrm{mod-2}}$. Obviously, to ensure Eq.~(\ref{creNew2D}), we only need to modify the $X^2$ terms of $\mathcal{L}^{\textrm{mod-1}}$ and $\mathcal{L}^{\textrm{mod-2}}$.

\subsubsection{Example I}
Adding an $X^2$ term to $\mathcal{L}^{\textrm{mod-1}}$, we obtain
\begin{eqnarray}
\mathcal{L}^{\textrm{ex1}}=\sum_{i=2,3,\cdots}f_i^{\textrm{ex1}}(V)(X+V)^i
+X-V,
\end{eqnarray} where $f_i^{\textrm{ex1}}(V)\geq 0$.
Obviously, this model satisfies the criterions:
\begin{eqnarray}
\mathcal{L}^{\textrm{ex1}}|=-2V,\quad \mathcal{L}^{\textrm{ex1}}_{,X}|=1.
\end{eqnarray}
So, it is one of the twinlike models of the canonical model. To upgrade this model to a special twinlike model, we require
\begin{eqnarray}
\mathcal{L}^{\textrm{ex1}}_{,XX}|=2f_2^{\textrm{ex1}}(V)=-\frac c{V}.
\end{eqnarray}
or,
\begin{eqnarray}
f_2^{\textrm{ex1}}(V)=-\frac c{2V}.
\end{eqnarray}
For our model,
\begin{eqnarray}
&&\left[ \mathcal{L}_{,XV}+2V(\mathcal{L}_{,XXX}-\mathcal{L}_{,XXV})\right] | = -\frac{6 c}{V}, \\
&&\left[ \mathcal{L}_{,VV}+\mathcal{L}_{,XV}+2V(\mathcal{L}_{,XVV}-\mathcal{L}_{,XXV}) \right] | =0, \\
&&(\mathcal{L}_{,V}+2V\mathcal{L}_{,XV})| =-1-4 c.
\end{eqnarray}
Obviously, identities~(\ref{crespe2}) and (\ref{crespe4}) are also violated by our model.

\subsubsection{Example II}
Now, let us turn to another model. We consider
\begin{eqnarray}
\mathcal{L}^{\textrm{ex2}}&=&1-\sqrt{1+2U}\sqrt{1-2X}
%\nonumber\\
+f_2^{\textrm{ex2}}(V)(X+V)^2.
\end{eqnarray}
For Eq.~(\ref{creNew2D}) to be true, we need
\begin{equation}
f_2^{\textrm{ex2}}(V)= -\frac{c+V+2 c V}{2 V+4 V^2}.
\end{equation}
As a consequence,
\begin{eqnarray}
&&\left[ \mathcal{L}_{,XV}+2V(\mathcal{L}_{,XXX}-\mathcal{L}_{,XXV})\right] | = -\frac{3 c}{V}, \\
&&\left[ \mathcal{L}_{,VV}+\mathcal{L}_{,XV}+2V(\mathcal{L}_{,XVV}-\mathcal{L}_{,XXV}) \right] | =0, \\
&&(\mathcal{L}_{,V}+2V\mathcal{L}_{,XV})| =-1-2 c.
\end{eqnarray}

So far, we have shown that Eqs.~(\ref{crespe1})-(\ref{crespe4}) are not necessary for the construction of special twinlike models. Only Eqs.~(\ref{ag1}), (\ref{ag2}) and (\ref{creNew2D}) are required. Now let us consider the construction of special twinlike models for a gravitational model.
\section{Special twinlike braneworld models}
\label{sec3}
Defect solutions can also be applied in higher dimensions. In 1983, Rubakov and Shaposhnikov considered the possibility that our world is a domain wall in a five-dimensional flat space-time~\cite{RubakovShaposhnikov1983}. The domain wall is generated by a background scalar field with a $\phi^4$ interaction. By introducing an Yukawa coupling between the Dirac field and the background scalar field, the authors of Ref.~\cite{RubakovShaposhnikov1983} found that massless left-chiral Dirac particle can be trapped on the wall. Later, it was found that gravity can also be localized on domain walls in five-dimensional warped space-times~\cite{RandallSundrum1999,RandallSundrum1999a}.

In this section, we consider the so-called thick braneworld models, which are extensions of the Rubakov-Shaposhnikov model in warped space-times~\cite{Gremm2000,DeWolfeFreedmanGubserKarch2000,CsakiErlichHollowoodShirman2000,Giovannini2001a,Giovannini2002,Giovannini2003} (for a review on thick brane, see~\cite{DzhunushalievFolomeevMinamitsuji2010}). There are some successful examples in twinlike thick brane models~\cite{BazeiaDantasGomesLosanoMenezes2011}, but no criteria or explicit examples of special twinlike thick brane models. Our aim of this section is to fill this blank.

The action of a lot of thick brane models can be written as follows:
\begin{eqnarray}
\label{actionGr}
S=\int d^5 x \sqrt{-g}\left(\frac{1}{2\kappa_5^2}R+\mathcal{L}(\phi,X)\right),
\end{eqnarray}
where $\kappa_5^2$ is the gravitational coupling, $g$ is the determinant of the metric $g_{MN}$, $R$ is the scalar curvature, and $\mathcal{L}(\phi,X)$ is the Lagrangian density of the background scalar field that generates the domain wall. In this section, $M, N=0,1,2,3,5$ represent the indices of bulk coordinates, and Greek letters $\mu, \nu=0,1,2,3$ denote brane coordinate indices.
For simplicity, let us call the extra dimension as $y=x^5$.

The kinetic term of the scalar field now becomes $X=-\frac12g^{MN}\nabla_M\phi\nabla_N\phi$. The standard braneworld model corresponds to the model with $\mathcal{L}=\mathcal{L}_0=X-V(\phi)$. As the previous section, $V$ is the self-interaction of the scalar field.
 The Einstein equations for action~(\ref{actionGr}) are
\begin{eqnarray}
G_{MN}\equiv R_{MN}- \frac{1}{2}g_{MN}R = \kappa _5^2{T_{MN}},
\end{eqnarray}
where the energy-momentum tensor is
\begin{eqnarray}
T_{MN}={g_{MN}}\mathcal{L} + \mathcal{L}_{,X}\nabla_M\phi \nabla_N\phi,
\end{eqnarray}
To get thick brane solutions, we choose the following metric~\cite{Gremm2000,CsakiErlichHollowoodShirman2000}
\begin{eqnarray}
\label{metric}
ds^2 = a^2(y)\eta_{\mu\nu}dx^\mu dx^\nu + dy^2,
\end{eqnarray}
where $\eta_{\mu\nu}=\textrm{diag}(-1,+1,+1,+1)$ is the four-dimensional Minkowski metric, and $a(y)=\textrm{e}^{A(y)}$ is called the warp factor.
We also assume that the scalar field is static, namely, $\phi=\phi(y)$. As a consequence, the energy density takes the following form:
\begin{eqnarray}
\rho=T_{00}=-\textrm{e}^{2A}\mathcal{L}.
\end{eqnarray}

With all these assumptions, we can now explicitly write the Einstein equations as follows
%\begin{subequations}\label{Eqy}
\begin{eqnarray}
\label{eqy1}
-3\partial_y^2 A& =& \kappa _5^2{\mathcal {L}_X}(\partial_y \phi)^2,\\
\label{eqy2}
6(\partial_y A)^2& =& \kappa _5^2({\cal L}+{\mathcal {L}_X}(\partial_y \phi)^2).
\end{eqnarray}
%\end{subequations}
The equation of motion for the scalar field is given by
\begin{eqnarray}
\label{eqscalar}
&&(\partial_y ^2\phi)(\mathcal{L}_{,X}+2X \mathcal{L}_{,XX})
+\mathcal{L}_{,\phi} %\nonumber\\
 - 2X\mathcal{L}_{,X\phi}=-4\mathcal{L}_{,X}(\partial_y \phi)( \partial_y A).
\end{eqnarray}
This equation can be derived from Eqs.~(\ref{eqy1}) and (\ref{eqy2}). Therefore, only two of the dynamical equations are independent.
For the case without gravity, Eq.~(\ref{eqscalar}) reduces to Eq.~(\ref{EoM}).

\subsection{The superpotential method}
To solve the Einstein equations, one can introduce the superpotential $W(\phi)$, such that
\begin{eqnarray}
\label{A0}
\partial_y A=-\frac{\kappa_5^2}{3}W(\phi).
\end{eqnarray}
Then for the standard model, we get
\begin{eqnarray}
\label{XW}
\partial_y \phi=W_\phi,\quad\textrm{or}\quad X=-\frac12 W_\phi^2,
\end{eqnarray}
from Eq.~(\ref{eqy1}), and
\begin{eqnarray}
\label{VW}
V=\frac12 W_\phi^2-\frac23\kappa_5^2 W^2.
\end{eqnarray}
from Eq.~(\ref{eqy2}).

Equations~(\ref{A0})-(\ref{VW}) constitute the first-order formalism of the canonical braneworld model~\cite{BazeiaBritoLosano2006}. This formalism reexpresses the original second-order Einstein equations to some first-order ones, which are easier to solve. With all these expressions, we know that the on-shell Lagrangian of the canonical model takes the form:
\begin{eqnarray}
\label{LW}
\mathcal{L}_0|_{X=-\frac12 W_\phi^2}=-W_\phi^2+\frac23\kappa_5^2 W^2.
\end{eqnarray}
In what follows the evaluation on-shell $|_{X=-\frac12 W_\phi^2}$ will be represented simply by $|$.

By definition, a twinlike braneworld model should share the same scalar field configuration, space-time geometry, and energy density with the canonical model~\cite{BazeiaDantasGomesLosanoMenezes2011,AdamQueiruga2011}.
The first two requirements can be fulfilled if the warp factor and the kinetic term $X$ of the noncanonical model also satisfy Eqs.~(\ref{A0}) and (\ref{XW}), respectively. Then, from Eq.~(\ref{eqy1}), one would obtain the following constraint~\cite{AdamQueiruga2011}:
\begin{eqnarray}
\label{twinBr1}
\mathcal{L}_{,X}|=1.
\end{eqnarray}
To fulfill the third requirement, the on-shell Lagrangian of the noncanonical model should be (see Eq.~(\ref{eqy2}))
\begin{eqnarray}
\label{twinBr2}
\mathcal{L}|=\frac23\kappa_5^2 W^2-W_\phi^2.
\end{eqnarray}
With Eqs.~(\ref{twinBr1})-(\ref{twinBr2}) we are ready to construct twinlike models for the canonical model. But to construct special twinlike models, we need to analyze the linear structure of the noncanonical models.

It is convenient for us analyze the linear fluctuation in the conformal coordinate $r$, which is defined by $dr=a^{-1} dy$. In the conformal coordinate, the metric reads
\begin{eqnarray}
ds^2 = a^2(r)(\eta_{\mu\nu}dx^\mu dx^\nu + dr^2).
\end{eqnarray}

\subsection{Linearization of noncanonical branes}
To linearize a noncanonical braneworld model, we need to consider the fluctuations around both the scalar and the metric
\begin{eqnarray}
 \phi &=&\bar{\phi}(r)+\delta\phi(x^P),\\
g_{MN}&=&\bar{g}_{MN}(r)+\delta{g}_{MN}(x^P).
\end{eqnarray}
It is more convenient to define $\delta g_{MN}\equiv a^2h_{MN}$.

It is always possible to decompose the metric perturbation into scalar, vector, and tensor components (see Ref.~\cite{ZhongLiu2013}):
\begin{eqnarray}
{h_{\mu r}} &=& {\partial _\mu }F + {G_\mu },\\
{h_{\mu \nu }} &=& {\eta _{\mu \nu }}A + {\partial _\mu }{\partial _\nu }B + 2{\partial _{(\mu }}{C_{\nu )}} + {D_{\mu \nu }},
\end{eqnarray}
where $C_\mu$ and $G_\mu$ are transverse vector perturbations:
\begin{eqnarray}
\partial^\mu C_\mu=0=\partial^\mu G_\mu,
\end{eqnarray}
and $D_{\mu \nu }$ is transverse and traceless (TT) tensor perturbation:
\begin{eqnarray}
\partial^\nu D_{\mu \nu }=0=D^\mu_\mu.
\end{eqnarray}
Note that the indices of the perturbations are always raised and lowered by $\eta^{\mu\nu}$, so that $\partial^{\mu}\equiv\eta^{\mu\nu}\partial_{\nu}$, and ${\square^{(4)}}\equiv\eta^{\mu\nu}\partial_\mu
\partial_\nu$.

The advantage of this decomposition is that different types of perturbations evolve independently. Therefore, the full linear spectrum of a braneworld model can be separated into scalar, vector, and tensor modes.

In Ref.~\cite{ZhongLiu2013}, we have systematically derived the quadratic action for all three types of fluctuation modes. So here we only briefly review the results. The quadratic action for the vector and the tensor modes are
\begin{eqnarray}
\label{37}
  {\delta ^{(2)}}{S_{\textrm{vector}}}& =& \frac{1}{2}\int d^4xdr {\hat{v}^\mu }{\square ^{(4)}}{\hat{v}_\mu },
\end{eqnarray}
and
\begin{eqnarray}
  &&{\delta ^{(2)}}{S_{\textrm{tensor}}}=\frac{1}{4}\int d^4xdr\hat{D}^{\mu \nu }
  \bigg\{\square ^{(4)}{\hat{D}_{\mu \nu }} + \hat{D}_{\mu \nu }'' - \frac{{({a^{\frac{3}{2}}})''}}{{{a^{\frac{3}{2}}}}}{\hat{D}^{\mu \nu }}\bigg\},
\end{eqnarray}
respectively, where
\begin{equation}
\hat{v}^\mu={a^{\frac{3}{2}}}({G_\mu } - C_\mu'),\quad {\hat{D}^{\mu \nu }} = {a^{\frac{3}{2}}}{D^{\mu \nu }},
\end{equation}
and primes represent the derivative with respect to $r$ in this section.

Obviously, the spectra of both the vector and tensor modes are independent of the Lagrangian of the noncanonical scalar field, they are determined only by the warp factor $a(r)$. Since twinlike models share the same geometry with the canonical braneworld model, they also share the same vector and tensor spectra with the canonical model.

It is the scalar modes which render the spectra of the twinlike models different. Thus, in order to construct special twinlike models, we need to find the condition under which the twinlike models also share the same scalar spectrum. The derivation of the quadratic action of the scalar modes is rather lengthy, we only cite the final result here (see Ref.~\cite{ZhongLiu2013} for details):
\begin{equation}
 {\delta ^{(2)}}{S_{\textrm{scalar}}} = \int d^4xdr^{\ast}\mathcal{G}
 \left\{
   {\square ^{(4)}}\mathcal{G}+ \ddot{\mathcal{G}} -\frac{\ddot{\theta}}{\theta}\mathcal{G}
\right\}.
\end{equation}
This action is similar to the one we obtained in Eq.~(\ref{G2}).
The new coordinate $r^{\ast}$ (corresponds to $x^\ast$), quantities $\theta$ and $\gamma$ are similarly defined as previous:
\begin{eqnarray}
dr^{\ast}&=&\gamma^{-1/2}dr,\\
\theta &=&\gamma^{1/4}z,\\
\gamma&=&1+2\frac{\mathcal{L}_{,XX} X}{\mathcal{L}_{,X}}.
\end{eqnarray}
The over dots on $\theta$ and $\mathcal{G}$ represent derivatives with respect to $r^\ast$.

What different is that now the normal mode of the scalar perturbations is defined by
\begin{equation}
\mathcal{G}=\frac{\kappa_5}{2}\gamma^{1/4}a^{3/2}\sqrt{\mathcal{L}_{,X}}
 \Big(2\delta \phi  - \frac{{\phi '}}{\mathcal{H}}A \Big),
\end{equation}
and that the quantity $z$ is given by
\begin{equation}
z=a^{3/2}\frac{\phi '}{\mathcal{H} }\sqrt{\mathcal{L}_{,X}},
\end{equation}
where $\mathcal{H}\equiv a'/a$.

Clearly, the linear spectrum of the normal mode $\mathcal{G}$ is determined only by $\ddot{\theta}/\theta$. For the canonical model $\mathcal{L}_{0,X}=1$ and $\mathcal{L}_{0,XX}=0$, we get
\begin{equation}
\theta_0=a^{3/2}\frac{\phi '}{\mathcal{H} }.
\end{equation}
For a twinlike model whose Lagrangian is already constrained by Eqs.~(\ref{twinBr1}) and (\ref{twinBr2}), the requirement that $\ddot{\theta}/\theta=\ddot{\theta}_0/\theta_0$ is equivalent to
\begin{equation}
(\mathcal{L}_{,XX} X)|=c,
\end{equation}
or
\begin{equation}
\label{specialBrane}
\mathcal{L}_{,XX}|=-\frac{2 c}{ W_\phi^2}.
\end{equation} We assume that $c$ is a positive constant.
Now we are ready to write some special twinlike braneworld models.

\subsection{Examples}
It is not necessary for us to start from zero. In fact, in Ref.~\cite{AdamQueiruga2011} the authors have constructed several twinlike braneworld models. Unfortunately, none of these braneworld models satisfies Eq.~(\ref{specialBrane}). Therefore, these models are not the special twinlike models of the canonical braneworld model. At the time when Ref.~\cite{AdamQueiruga2011} was written, the structure of linear scalar perturbation of an arbitrary $K$-field braneworld model was still an open question. So the authors of Ref.~\cite{AdamQueiruga2011} did not address when two twinlike braneworld models would also possess an identical linear structure.

As shown in the previous subsection, the effective potential for the scalar linear perturbation is determined only by $\mathcal{L}_{,XX}$. So, we only need to modify the $X^2$ terms of the models given by Ref.~\cite{AdamQueiruga2011} to make them satisfy Eq.~(\ref{specialBrane}).

For the first model, we consider
\begin{eqnarray}
\mathcal{L}^{\textrm{ex1}}=f^{\textrm{ex1}}(\phi)\Big(X+\frac12 W_\phi^2\Big)^2
+X-V,
\end{eqnarray} where $f^{\textrm{ex1}}(\phi)\geq 0$, and $V$ is given by Eq.~(\ref{VW}). Obviously, this Lagrangian satisfies Eqs.~(\ref{twinBr1}) and (\ref{twinBr2}). In order to satisfy Eq.~(\ref{specialBrane}), $f(\phi)$ should take the following form:
\begin{eqnarray}
f^{\textrm{ex1}}(\phi)=-\frac c{W_\phi^2}.
\end{eqnarray}

For the second model, we consider the DBI type model:
\begin{eqnarray}
\mathcal{L}^{\textrm{ex2}}&=&1-\sqrt{1+W_\phi^2}\sqrt{1-2X} %\nonumber\\
+\frac23\kappa_5^2 W^2+f^{\textrm{ex2}}(\phi)\Big(X+\frac12 W_\phi^2\Big)^2.
\end{eqnarray}
This model will be a special twinlike braneworld model if
\begin{eqnarray}
f^{\textrm{ex2}}(\phi)=-\frac{c}{W_{\phi }^2}-\frac{1}{2 \left(1+W_{\phi }^2\right)}.
\end{eqnarray}
The above examples show that it is possible to construct special twinlike models in gravitational systems, in which $\mathcal{L}_{,XX}$ can be a function of the extra dimension. For comparison and simplicity, we only studied the two types of models given by Ref.~\cite{AdamQueiruga2011}. But in fact, one can construct infinite special twinlike models for the canonical braneworld model. Besides, although we only considered braneworld models, it is not difficult to repeat the same procedures for cosmological models.

\section{Summary and comments}
\label{sec4}
Field configuration, energy density, and linear spectrum are important features of a defect solution. Twinlike defects are defect solutions that share the same field configuration and energy density. Usually, they can be distinguished by their linear spectra. However, there are some special cases, where the twinlike defects even share the same linear spectrum. We call such special defects the special twinlike defects. Some special twinlike defects have been constructed in Refs.~\cite{BazeiaMenezes2011,AdamQueiruga2012}. Especially, the authors of Ref.~\cite{AdamQueiruga2012} derived the criteria for special twinlike defects in two-dimensional flat space-time. They argue that the Lagrangian of a special twinlike model should satisfy the on-shell condition $\mathcal{L}_{,XX}|=0$. It is interesting to extend the works of Refs.~\cite{BazeiaMenezes2011,AdamQueiruga2012} to curved and higher space-time and to cases with more general on-shell condition where $\mathcal{L}_{,XX}|\neq 0$.

In this paper, we successfully constructed special twinlike models in both flat and warped space-times. We showed that in both cases it is possible to construct special twinlike models with $\mathcal{L}_{,XX}|\neq0$. The criteria (\ref{creNew2D}) and (\ref{specialBrane}) does not depend on explicit solutions. Because the solution is determined only by the superpotential $W(\phi)$: given a superpotential, we can find the corresponding solution of $\phi(y)$ and $a(y)$. But in our discussions above, we did not specify the form of $W$. Thus our results are valid for all the special twinlike models \emph{correspond to the canonical model} $\mathcal{L}_0$. Of course, it is also interesting to consider special twinlike models \emph{correspond to noncanonical models}, in this case our criteria will be modified, however.

The existence of the special twinlike models makes it a theoretical problem to distinguish the noncanonical models from the canonical one. Because for any phenomenologically acceptable canonical scalar field model, we can always construct infinite special twinlike models, which share the same background behaviors and linear structure with the canonical model.

As the normal twinlike models, the special twinlike models are not a reparametrization of the canonical model in general. One of the possible ways to distinguish the canonical model from its special twins is to consider perturbations beyond the linear order. As the higher-order perturbations are considered, the present work might be interested. Because in principle, models with $\mathcal{L}_{,XX}|\neq0$ are different from the canonical model (where $\mathcal{L}_{,XX}|=0$). For sure, models with nontrivial $\mathcal{L}_{,XX}|$ would have a richer nonlinear structure than those with $\mathcal{L}_{,XX}|=0$. It has been shown, at least in the frame of cosmology~\cite{ChenHuangKachruShiu2007}, that the third-order Lagrangian of the scalar perturbation is determined by both $\mathcal{L}_{,XX}$ and $\mathcal{L}_{,XXX}$. It is known that the nonlinear perturbation is one of the candidates to offer the primordial non-Gaussianity in the microwave background radiation.
Similarly, $\mathcal{L}_{,XX}$ might play an important role in other noncanonical scalar field models. For this reason, special twinlike models with nontrivial $\mathcal{L}_{,XX}|$ deserve further studies.

Another possible way to distinguish two special twinlike models is to study the quantum effects. The quantization of a space-dependent static defect configuration $\phi_c(x)$ can be realized by expanding the action around $\phi_c(x)$, and treating the fluctuation $\delta\phi(x,t)=\phi(x,t)-\phi_c(x)$ as quantum operator (see \cite{Rajaraman1982} for a pedagogical introduction). To the lowest order, we will obtain the quadratic action for $\delta\phi$, which is nothing but Eq.~(\ref{G2}). Depending on the form of $\phi_c(x)$, the spectrum of Eq.~(\ref{G2}) might consist a zero mode, a few bound states and a continuum of scattering states~\cite{Rajaraman1982}. As the massive modes are considered, no matter they are bound states or scattering states, the special twinlike defects cannot be distinguished. Because as we have shown in the subsection~\ref{secQuadratic}, the special twinlike defects share the same linear perturbation equation. However, one should pay a special attention to the zero mode, whose quantization is achieved by introducing the collective coordinate or the modulous\footnote{In this paper, we only considered co-dimension one kink solutions. Only one spatial translational symmetry is broken by the kink solution, so there is a single collective coordinate, namely, the location of the kink. In models with more translational and inner symmetries, there are more collective coordinates which constitute the moduli space~\cite{Tong2005}.}~\cite{ChristLee1975,Tomboulis1975}. It is possible that special twinlike models can be distinguished even at the lowest order, due to the quantization of the zero mode. A proof to this conjecture is beyond the scope of the present work, we would leave it to our future work.

\section*{Acknowledgments}

We thank the anonymous referees for all their kind criticisms and helpful suggestions. This work was supported by the National Natural Science Foundation of China (Grant No. 11375075), and the Fundamental Research Funds for the Central Universities (Grants No. lzujbky-2015-jl1). Y.Z. was also supported by the scholarship granted by the Chinese Scholarship Council (CSC).

\section*{References}
\providecommand{\newblock}{}

%\bibliographystyle{iopart-num}
%\bibliography{D:/jabref/library/articles/articles}

\end{document}